\documentclass{aa}
\usepackage{natbib}
\usepackage[dvips,final]{graphics}
\usepackage{fix2col}
\usepackage{dcolumn}
\usepackage{psfrag}

\psfrag{l}{$\lambda_{0}$}
\psfrag{W}{$\Omega_{0}$}

\begin{document}

\thesaurus{02(12.07.1; 12.03.1; 12.03.4; 12.03.3)}

\title{
Gravitational lensing statistics with extragalactic surveys
}

\subtitle{
IV. Joint constraints from gravitational lensing statistics and CMB 
anisotropies
}

\author{
Juan Francisco Macias-Perez\inst{1}
\and
Phillip Helbig\inst{1,2}
\and
Ralf Quast\inst{3} 
\and
Althea Wilkinson\inst{1} 
\and
Rod Davies\inst{1}
} 

\institute{
University of Manchester, 
Jodrell Bank Observatory,
Macclesfield,
Cheshire SK11 9DL, 
UK
\and 
University of Groningen,
Kapteyn Astrononomical Institute,
P.~O. Box 800,
9700 AV Groningen,
The Netherlands
\and
University of Hamburg, 
Hamburg Observatory, 
Gojenbergsweg 112, 
21029 Hamburg, 
Germany
}

\date{03 June 1999 / Accepted 21 October 1999}

\offprints{P.~Helbig}
\mail{p.helbig@jb.man.ac.uk}

\authorrunning{J.~F.~Macias-Perez et al.}
\titlerunning{
Gravitational lensing statistics with extragalactic surveys.  IV
}

\maketitle

\begin{abstract}
We present constraints on the cosmological constant $\lambda_{0}$ and
the density parameter $\Omega_{0}$ from joint constraints from the
analyses of gravitational lensing statistics of the Jodrell Bank-VLA
Astrometric Survey (JVAS), optical gravitational lens surveys from the
literature and CMB anisotropies.  This is the first time that
quantitative joint constraints involving lensing statistics and CMB
anisotropies have been presented.  Within the assumptions made, we
achieve very tight constraints on both $\lambda_{0}$ and $\Omega_{0}$. 
These assumptions are cold dark matter models, no tensor components, no
reionisation, CMB temperature $T_{\mathrm{CMB}}=2.728$, number of
neutrinos $n_{\nu}=3$, helium abundance $Y_{\mathrm{He}}=0.246$,
spectral index $n_{\mathrm{s}}=1.0$, Hubble constant
$H_{0}=68$\,km\,s$^{-1}$\,Mpc$^{-1}$, baryonic density
$\Omega_{\mathrm{b}}=0.05$.  All models were normalised to the
\textit{COBE} data and no closed models ($k=+1$) were computed. 
Using the CMB data alone, the best-fit model has $\lambda_{0}=0.60$ and
$\Omega_{0}=0.34$ and at 99\% confidence the lower limit on
$\lambda_{0}+\Omega_{0}$ is $0.8$.  Including constraints from
gravitational lensing statistics doesn't change this significantly, 
although it does change the allowed region of parameter space. 
A universe with
$\lambda_{0}=0$ is ruled out for any value of $\Omega_{0}$ at better
than 99\% confidence using the CMB alone. Combined with constraints from
lensing statistics, $\lambda_{0}=0$ is also ruled out at better than 
99\% confidence. 

As the region of parameter space allowed by the CMB is, within our 
assumptions, much smaller than that allowed by lensing statistics, the 
main result of combining the two is to change the range of parameter 
space allowed by the CMB along its axis of degeneracy.
\keywords{
gravitational lensing -- cosmic microwave background -- cosmology: theory -- 
cosmology: observations 
}
\end{abstract}

%%%%%%%%%%%%%%%%%%%%%%%%%%%%%%%%%%%%%%%%%%%%%%%%%%%%%%%%%%%%%%%%%%%%%%%%
\section{Introduction}

Cosmological tests which are sensitive to $\lambda_{0}$ and $\Omega_{0}$
(the normalised cosmological constant and density parameter,
respectively) can be used to construct likelihood contours in the
$\lambda_{0}$-$\Omega_{0}$ plane.  Each test usually has a degeneracy
such that moving along a curve (which often approximates a line) in the
$\lambda_{0}$-$\Omega_{0}$ plane leaves the likelihood (almost)
unchanged.  It has long been realised 
\citep[e.g.][]{DEisensteinHT98b,DEisensteinHT99}
that the direction of degeneracy of constraints from cosmic microwave
background anisotropies is roughly orthogonal to that of most other
tests.  Thus, combining the constraints from CMB anisotropies with those
from other cosmological tests can give much tighter constraints than
either alone. 

Gravitational lensing statistics provide constraints which are
degenerate such that $\lambda_{0}$ and $\Omega_{0}$ values are
positively correlated.  This is also the case with cosmological tests
such as the product of the Hubble constant and the age of the universe,
the angular size-redshift (standard rod) test and the
luminosity-redshift (standard candle) test. The opposite is the case
with constraints derived from CMB anisotropies.  Thus, it seems natural
to combine the constraints from \citet[hereafter
Paper~I]{RQuastPHelbig99a} and \citet[hereafter
Paper~II]{PHelbigMQWBK99a} with an analysis of the type performed by
\citet[hereafter L98]{CLineweaver98a}, which in itself already provides
quite tight constraints. 

It is important to note that all three of our analyses have fixed all
parameters except $\lambda_{0}$ and $\Omega_{0}$ (though in Papers~I and
II an attempt has been made to estimate the effect of the uncertainty of
the other parameters on the derived constraints on $\lambda_{0}$ and
$\Omega_{0}$ by varying one parameter by two standard deviations (see
Paper~I)). Ideally, an investigation such as the present one should
incorporate the uncertainties in all input parameters into the analysis.
Such a programme is currently under development. 

In this work, we use the most recent CMB data available to do an
analysis similar to that of L98 and combine the constraints with the
lensing statistics constraints from Papers~I and II following the
procedure outlined in \citet[hereafter Paper~III]{PHelbig99a}.  The
plan of this paper is as follows.  In Sect.~\ref{cmb} we describe the
procedure used to calculate likelihoods in the
$\lambda_{0}$-$\Omega_{0}$ plane from CMB data and in
Sect.~\ref{results} we discuss our results and compare them with those
from Papers~I and II.  Sect.~\ref{conclusions} summarises our
conclusions.  For a comparison with other recent constraints from a
variety of cosmological tests, see Paper~I. 

Throughout, as in Papers~I, II and III, $\lambda_{0}=\Lambda/3H_{0}^{2}$
and $\Omega_{0}=8\pi G\rho_{0}/3H_{0}^{2}$ refers to the density of
matter, i.e.~not some `total density' which in our notation would be
$\lambda_{0}$+$\Omega_{0}$ (or perhaps including a contribution from
pressure as well, which we consider to be irrelevant here).  The index
$0$ refers to present-day values, since in general these quantities are
time-dependent.  (See Paper~I for an overview of the rest of our
notation and general description of the gravitational lensing statistics
method.)

%%%%%%%%%%%%%%%%%%%%%%%%%%%%%%%%%%%%%%%%%%%%%%%%%%%%%%%%%%%%%%%%%%%%%%%%
\section{Constraints from CMB anisotropies}
\label{cmb}

Gravitational lensing statistics are not very sensitive to $\Omega_{0}$.
However, CMB data can constrain $\Omega_{0}$ more effectively (L98).
Following \citet[hereafter L97]{CLineweaverBBB97a} we have calculated
probability contours in the $\lambda_{0}$-$\Omega_{0}$ plane. This
method is based on a $\chi^{2}$ minimisation: \begin{equation}  
\chi^{2}(\lambda_{0,i},\Omega_{0,i})=\sum_{i=1}^{N_{\mathrm{exp}}}  
\frac{(\mathrm{model}_{i}(\lambda_{0,i},\Omega_{0,i}) -  
\mathrm{temp}_{i})^{2}}{\sigma_{i}^{2}} \end{equation} where
$N_{\mathrm{exp}}$ is the number of experiments, $\mathrm{model}_{i}$ is
the theoretical predicted fluctuation at the multipole range covered by
the $i$-th experiment and $\mathrm{temp}_{i}$ represents the sky
fluctuation temperature measured by the $i$-th CMB experiment.  Each
pair $\lambda_{0,i}$,$\Omega_{0,i}$ in the $\lambda_{0}$-$\Omega_{0}$
plane corresponds to a model.  We constructed a matrix of models and
calculated the $\chi^{2}$ and the likelihood associated with it,
$\mathrm{e}^{-\frac{\chi^{2}}{2}}$.  The theoretical power spectra were
calculated with the help of
CMBFAST\footnote{\texttt{http://www.sns.ias.edu/\symbol{126}matiasz/CMBFAST/cmbfast.html}}
\citep{MZaldarriaga98a}. The models depend on a range of parameters.  To
make the test computationally feasible, we fixed all of them except
$\Omega_{0}$ and $\lambda_{0}$.  We consider cold dark matter models, no
tensor components and no reionisation.  No closed models ($k=+1$) were
computed because CMBFAST does not yet support this (we are looking
forward to the new CMBFAST version which will include these models). The
CMB temperature was set to $T_{\mathrm{CMB}}=2.728$, the number of
neutrinos to $n_{\nu}=3$ and the helium abundance to
$Y_{\mathrm{He}}=0.246$. The spectral index used was
$n_{\mathrm{s}}=1.0$. All models were normalised to the \textit{COBE}
data.\footnote{See the last paragraph of this section.}
Finally, the Hubble constant and the baryonic density were set to
$H_{0}=68$\,km\,s$^{-1}$\,Mpc$^{-1}$ and $\Omega_{\mathrm{b}}=0.05$. All
these values were based on the best literature estimates and on the L98
conclusions.  $\lambda_{0}$ and $\Omega_{0}$ vary in the range $-0.48
\leq \lambda_{0} \leq 1.48$ and $0.06\leq\Omega_{0}\leq 0.98$ with a
resolution of $0.04$.  $\Omega_{0}=0.02$ models were not computed since
this is inconsistent with our value for $\Omega_{\mathrm{b}}$; these 
models, and all outside the examined parameter space, were assigned an 
a priori likelihood of zero.  Otherwise, we have used a uniform prior. 
(See Paper~I for further discussion.)
(Note that
this is smaller than the range of parameter space covered in Papers~I
and II, but with a finer resolution.  Initially, we explored the 
parameter space as follows: $\lambda_{0}$ and $\Omega_{0}$ vary in the 
range $-3\leq \lambda_{0} \leq 1$ and $0.1\leq\Omega_{0}\leq 1.5$ with a
resolution of $0.1$, slightly smaller than the range
of parameter space covered in Papers~I and II but with the same 
resolution.  We restricted ourself to the higher resolution calculations 
in the smaller area of parameters space as outside of this no 
significant likelihood is present.) 

To compare data and models, the models have to be convolved with the
window function of the CMB experiments.  The window function delimits
the multipole range to which the experiment is sensitive.  This can be
seen as a Fourier transform of the experimental beam function in the
multipole space \citep{MWhiteMSrednicki95a}. On the other hand, to
compare results from different CMB experiments, the effect of the window
function must be removed.  This is accomplished by deconvolving both the
model and the data (see L97 for more details) . The quantities
$\mathrm{model}_{i}$ and $\mathrm{temp}_{i}$ are the deconvolved values.
 
We have built up a CMB data compilation that is based on L98 and on the
web page provided by 
Tegmark.\footnote{\texttt{http://www.sns.ias.edu/\symbol{126}max/cmb/experiments.html}}
We have also added new data from the Tenerife radiometers and
interferometer \citep{SDickerMDGRHDWHW99a}.  A list of the data used with
their references can be found in Table~\ref{ta:data}. The window
functions of each of the experiments have been also gathered.  We have
calculated some of them from analytical expressions
\citep{MWhiteMSrednicki95a}. The rest can be found in each of the CMB
experiment web pages which can be accessed from Tegmark's web page
mentioned at the start of this paragraph. 
%%%%%%%%%%%%%%%%%%%%%%%%%%%%%%%%%%%%%%%%%%%%%%%%%%%%%%%%%%%%%%%%%%%%%%%%
%%%%%%%%%%%%%%%%%%%%%%%%%%%%%%%%%%%%%%%%%%%%%%%%%%%%%%%%%%%%%%%%%%%%%%%%
\begin{table*}
\caption[]{CMB data used.  The window function is centered at
$l=l_{\mathrm{eff}}$ and drops to half of its central value at
$l_{\mathrm{min}}$ and $l_{\mathrm{max}}$, except for COBE, where
$l_{\mathrm{min}}$ and $l_{\mathrm{max}}$ instead indicate the RMS width
of the window function.  The COBE points from \citet{MTegmarkAHamilton97a} 
are not actually used in our $\chi^{2}$ analysis, but are included here 
since they appear in Fig.~\ref{fi:bestfit}; see text for details}  
\label{ta:data}
\newcolumntype{d}{D{.}{.}{-1}}
\begin{tabular*}{\linewidth}{@{\extracolsep{\fill}}lddddddl}
\hline
Experiment   
& \multicolumn{1}{c}{$\delta T$  ($\mu$K)}
& \multicolumn{1}{c}{$+$ ($\mu$K)}
& \multicolumn{1}{c}{$-$ ($\mu$K)}
& \multicolumn{1}{c}{$l_{\mathrm{min}}$}
& \multicolumn{1}{c}{$l_{\mathrm{eff}}$}
& \multicolumn{1}{c}{$l_{\mathrm{max}}$}
& Reference \\
\hline                                                      
COBE 1       & 8.5   & 16.0  & 8.5  & 2    & 2.1  & 2.5  & \citet{MTegmarkAHamilton97a} \\
COBE 2       & 28.0  & 7.4   & 10.4 & 2.5  & 3.1  & 3.7  & \citet{MTegmarkAHamilton97a} \\
COBE 3       & 34.0  & 5.9   & 7.2  & 3.4  & 4.1  & 4.8  & \citet{MTegmarkAHamilton97a} \\
COBE 4       & 25.1  & 5.2   & 6.6  & 4.7  & 5.6  & 6.6  & \citet{MTegmarkAHamilton97a} \\
COBE 5       & 29.4  & 3.6   & 4.1  & 6.8  & 8.0  & 9.3  & \citet{MTegmarkAHamilton97a} \\
COBE 6       & 27.7  & 3.9   & 4.5  & 9.7  & 10.9 & 12.2 & \citet{MTegmarkAHamilton97a} \\
COBE 7       & 26.1  & 4.4   & 5.3  & 12.8 & 14.3 & 15.7 & \citet{MTegmarkAHamilton97a} \\
COBE 8       & 33.0  & 4.6   & 5.4  & 16.6 & 19.4 & 22.1 & \citet{MTegmarkAHamilton97a} \\
FIRS         & 29.4  & 7.8   & 7.7  & 3.0  & 10   & 30.0 & \citet{KGangaPCM94a} \\
Tenerife     & 32.5  & 10.1  & 8.5  & 13   & 20   & 31   & \citet{SHancockGDLRRWT97a} \\
SP           & 32.21 & 7.44  & 4.08 & 31   & 57   & 106  & \citet{JGundersonLSWFGKMSCSL95a} \\ 
BAM          & 55.6  & 27.4  & 9.8  & 28   & 74   & 97   & \citet{STuckerGHST97a}  \\
ARGO         & 42.01 & 6.41  & 7.06 & 52   & 95   & 176  & \citet{PdBdGMV} \\
MAX          & 43.44 & 7.24  & 4.94 & 78   & 145  & 263  & \citet{STanakaCDFGHHLLLMRSS96a} \\
Python 1     & 54.0  & 14.0  & 12.0 & 68   & 92   & 129  & \citet{SPlattKDPR96a} \\
Python 2     & 58.0  & 15.0  & 13.0 & 119  & 177  & 243  & \citet{SPlattKDPR96a} \\ 
IAC/Bartol   & 55.0  & 27.0  & 22.0 & 35   & 53   & 79   & \citet{BFemeniaRGLP98a} \\ 
MSAM1        & 48.42 & 11.95 & 7.95 & 86   & 160  & 251  & \citet{EChengCFIKMPPRS96a} \\
MSAM2        & 59.34 & 12.08 & 8.23 & 173  & 263  & 383  & \citet{EChengCFIKMPPRS96a} \\
QMAP F1 Ka   & 49.0  & 6.0   & 7.0  & 47   & 92   & 157  & \citet{MDevlindOCHMNPT98a} \\
QMAP F1 Q    & 47.0  & 8.0   & 10.0 & 38   & 84   & 140  & \citet{MDevlindOCHMNPT98a} \\
QMAP F2 Ka   & 46.0  & 10.0  & 12.0 & 44   & 91   & 138  & \citet{THerbigAdOCDMPM98a} \\
QMAP F2 Ka   & 63.0  & 10.0  & 12.0 & 81   & 145  & 209  & \citet{THerbigAdOCDMPM98a} \\
QMAP F2 Q    & 56.0  & 5.0   & 6.0  & 58   & 125  & 192  & \citet{THerbigAdOCDMPM98a} \\
QMAP F1+2 Ka & 47.0  & 6.0   & 7.0  & 39   & 80   & 121  &\citet{dOCDHMBNP98a} \\
QMAP F1+2 Ka & 59.0  & 6.0   & 7.0  & 72   & 126  & 180  &\citet{dOCDHMBNP98a} \\
QMAP F1+2 Q  & 52.0  & 5.0   & 5.0  & 47   & 111  & 175  & \citet{dOCDHMBNP98a} \\
Saskatoon 1  & 49.0  & 8.0   & 5.0  & 53   & 86   & 132  & \citet{CNetterfieldDJPW97a} \\
Saskatoon 2  & 69.0  & 7.0   & 6.0  & 119  & 166  & 206  & \citet{CNetterfieldDJPW97a} \\
Saskatoon 3  & 85.0  & 10.0  & 8.0  & 190  & 236  & 274  & \citet{CNetterfieldDJPW97a} \\
Saskatoon 4  & 86.0  & 12.0  & 10.0 & 243  & 285  & 320  & \citet{CNetterfieldDJPW97a} \\
Saskatoon 5  & 69.0  & 19.0  & 28.0 & 304  & 348  & 401  & \citet{CNetterfieldDJPW97a} \\
CAT 1        & 48.44 & 7.67  & 5.71 & 339  & 422  & 483  & \citet{PScottSPOSLJHOSB96a} \\
CAT 2        & 45.20 & 11.02 & 8.13 & 546  & 610  & 722  & \citet{PScottSPOSLJHOSB96a} \\
RING5M 2     & 56.0  & 8.5   & 6.6  & 361  & 589  & 756  & \citet{ELeitchRPMG98a} \\ 
\hline
\end{tabular*}
\end{table*}
%%%%%%%%%%%%%%%%%%%%%%%%%%%%%%%%%%%%%%%%%%%%%%%%%%%%%%%%%%%%%%%%%%%%%%%%

We do not actually use the COBE points from
\citet{MTegmarkAHamilton97a}, since the COBE data are used internally by
CMBFAST.  We include them in Table~\ref{ta:data} since they appear in
Fig.~\ref{fi:bestfit}.  On the one hand, CMBFAST normalises the power
spectra to COBE according to the fitting formula given in
\citet{EBunnMWhite97a}.  In order to take into account the \emph{shape}
of the power spectrum in the region of the COBE data, as well as its
\emph{amplitude}, we have multiplied the likelihood obtained from our
$\chi^{2}$ analysis (without the COBE points) with the likelihood (again
provided by CMBFAST using a formula from \citet{EBunnMWhite97a}) of the
corresponding power spectra relative to a flat power spectrum.

%%%%%%%%%%%%%%%%%%%%%%%%%%%%%%%%%%%%%%%%%%%%%%%%%%%%%%%%%%%%%%%%%%%%%%%%
\section{Results}
\label{results}

\subsection{CMB results}

\begin{figure}
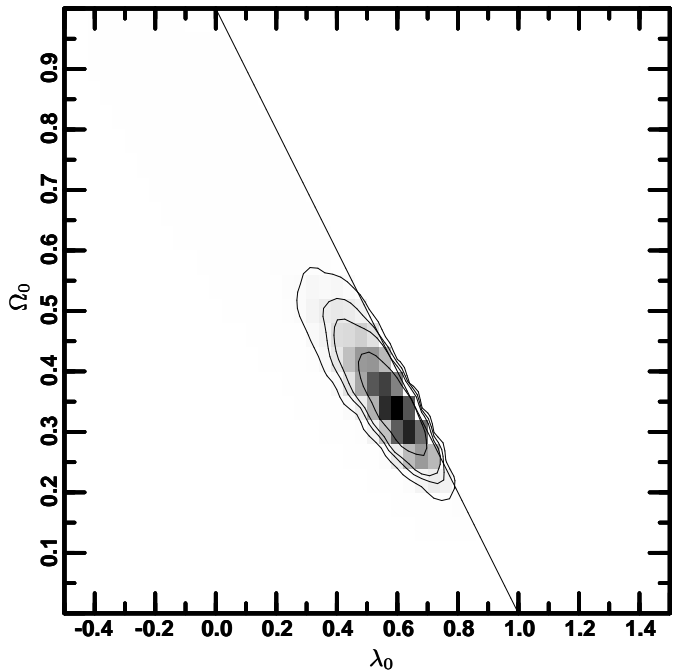

\resizebox{\columnwidth}{!}{\includegraphics{8913.f01}}%
\makebox[0pt][r]{\resizebox{\columnwidth}{!}{\includegraphics{8913.f00}}}
\caption[]{
The likelihood function $p(D|\lambda_0,\Omega_0,\vec{\xi}_0)$ based on
the CMB data in Table \ref{ta:data}.  ($\vec{\xi}_0$ represents the 
parameters other than $\lambda_{0}$ and $\Omega_{0}$; $\vec{\xi}_0$, 
which corresponds to `nuisance parameters', was 
held constant for all calculations in this paper.
See \citet[Paper~I]{RQuastPHelbig99a} for definitions and further discussion.) 
All nuisance parameters are assumed to precisely take their mean values.
The pixel grey level is directly proportional to the likelihood ratio,
darker pixels reflect higher ratios.  In this and all subsequent plots, 
unless noted otherwise, the pixel size reflects the
resolution of our numerical computations, the contours mark the
boundaries of the minimum $0.68$, $0.90$, $0.95$ and $0.99$ confidence
regions for the parameters $\lambda_0$ and $\Omega_0$ and are `real 
contours' in the sense of the discussion in 
\citet[Paper~III]{PHelbig99a}.  The diagonal line corresponds to 
$k=0$; the area to the right of this corresponds to spatially closed 
models which were not examined here.  The fact that some grey pixels and 
contours are nevertheless in this region is due to finite resolution and 
interpolation, respectively}
\label{fi:cblik}
\end{figure}

Fig.~\ref{fi:cblik} shows the likelihood 
$\mathrm{e}^{-\frac{\chi^{2}}{2}}$ obtained from the $\chi^{2}$
calculations over our matrix of cosmological models, for the CMB data.
The contours
correspond to 68\%, 90\%, 95\% and 99\% confidence levels, i.e.~the
area within the x\% contour level contains x\% of the sum of all the
likelihood values (one per pixel) in the plot. (See Paper~III for 
further discussion of this point, which is important for the detailed 
comparison of different results in the literature.)  
%%%%%%%%%%%%%%%%%%%%%%%%%%%%%%%%%%%%%%%%%%%%%%%%%%%%%%%%%%%%%%%%%%%%%%
%%%%%%%%%%%%%%%%%%%%%%%%%%%%%%%%%%%%%%%%%%%%%%%%%%%%%%%%%%%%%%%%%%%%%%
%%%%%%%%%%%%%%%%%%%%%%%%%%%%%%%%%%%%%%%%%%%%%%%%%%%%%%%%%%%%%%%%%%%%%%%%%%%
These confidence limits differ from those used in L98 in two ways.
First, we plot the x\% contour as that which encloses x\% of the
integrated probability density in the $\lambda_{0}$-$\Omega_{0}$ plane,
i.e.~joint probability contours in $\lambda_{0}$ and $\Omega_{0}$,
whereas those in L98 correspond to the appropriate confidence levels
\emph{when projected onto one of the axes}.  Thus, our contours are
naturally larger than those of L98.  Second, the contours of L98 are
actually $\Delta\chi^{2}$ contours, which correspond to the appropriate
confidence intervals if Gaussianity is assumed, whereas ours are `real'
confidence contours as defined above. (See Papers~I and III for further
discussion.) 

It is well known that the errors quoted for CMB temperature fluctuation
measurements are not Gaussian.  Actually, for most of the experiments the
error bars quoted are asymmetric.  The observational temperature
fluctuations are usually calculated using maximum likelihood techniques,
with the value corresponding to the mean and the error to the 1-$\sigma$
cutoff.  We have taken asymmetric error bars into account by using the
positive or negative error bar according to whether the theoretical
value falls above or below the experimental value.  There is perhaps some
disagreement as to the errors that the assumption of Gaussianity
introduces on the constraints on $\lambda_{0}$ and $\Omega_{0}$. The L98
confident limits are based on the $\chi^2$ method, which assumes that
the likelihood function is a Gaussian.  Other groups
\citep[e.g.][]{JBartlettBLDD98,JBartlettBDLD98a,JBartlettBDLD98b,JBartlettDBLD99a} 
consider this
approximation to be not justified and base their calculations on
\emph{likelihood functions}.  Unfortunately, likelihood functions are
not provided for all the CMB experiments and their calculations are
based on approximations.  How to compute this approximation is still
poorly understood.  We have opted for a much simpler approach by using
the likelihood defined above instead of the confidence limits provided
by the $\chi^{2}$ statistics.  Contours using the maximum likelihood
approach seem to be larger than those from Gaussian statistics.  Our
contours are found to lie between the approach of L97 and L98 and that
of \citet{JBartlettBDLD98a,JBartlettBDLD98b}.\footnote{`Gaussianity' is
an issue in at least three different contexts with relation to
cosmological constraints derived from CMB anisotropies.  First, the
correspondence between $\Delta\chi^{2}$ values or fractions of the peak
likelihookd and `real' confidence contours as defined above often
assumes Gaussianity.  Second, not unrelated, there is the issue of the
Gaussianity of the error bars of individual experiments.  Third is the
question whether the primordial density fluctuations are Gaussian.}

\subsection{Review of lens statistics results}

\begin{figure*}
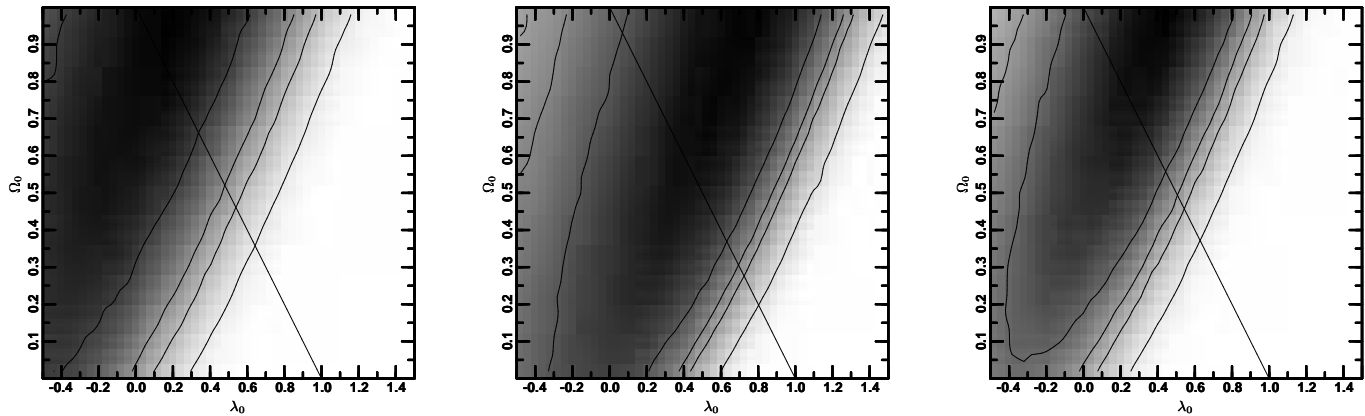

\noindent
\resizebox{0.3\textwidth}{!}{\includegraphics{8913.f02}}%
\makebox[0pt][r]{\resizebox{0.3\textwidth}{!}{\includegraphics{8913.f00}}}
\hfill
\resizebox{0.3\textwidth}{!}{\includegraphics{8913.f03}}%
\makebox[0pt][r]{\resizebox{0.3\textwidth}{!}{\includegraphics{8913.f00}}}
\hfill
\resizebox{0.3\textwidth}{!}{\includegraphics{8913.f04}}%
\makebox[0pt][r]{\resizebox{0.3\textwidth}{!}{\includegraphics{8913.f00}}}
\caption[]{
\emph{Left:}
The likelihood function $p(D|\lambda_{0},\Omega_{0})$ from
optical gravitational lens surveys discussed in
\citet[Paper~I]{RQuastPHelbig99a}, but with a higher resolution and 
confined to a smaller area of parameter space.  (This makes the 
positions of the contours slightly different; see 
\citet[Paper~III]{PHelbig99a} for a discussion.)
\emph{Centre:}
The same but from the analysis of JVAS presented in
\citet[Paper~II]{PHelbigMQWBK99a}.
\emph{Right:}
The same but joint constraints from the two other plots as discussed in 
Paper~II}
\label{fi:lenses}
\end{figure*}

We have repeated the lens statistics calculations of Papers~I and II 
at the higher resolution ($\Delta\lambda_{0}=\Delta\Omega_{0}=0.04$) and
in the area ($-0.48 \leq \lambda_{0} \leq 1.48$ and
$0.02\leq\Omega_{0}\leq 0.98$) used in the CMB calculations.  For
comparison with Papers~I and II, these are shown in
Fig.~\ref{fi:lenses}. 
%%%%%%%%%%%%%%%%%%%%%%%%%%%%%%%%%%%%%%%%%%%%%%%%%%%%%%%%%%%%%%%%%%%%%%%%
%%%%%%%%%%%%%%%%%%%%%%%%%%%%%%%%%%%%%%%%%%%%%%%%%%%%%%%%%%%%%%%%%%%%%%%%
%\suppressfloats
%%%%%%%%%%%%%%%%%%%%%%%%%%%%%%%%%%%%%%%%%%%%%%%%%%%%%%%%%%%%%%%%%%%%%%%%

\subsection{Joint constraints}

\begin{figure*}
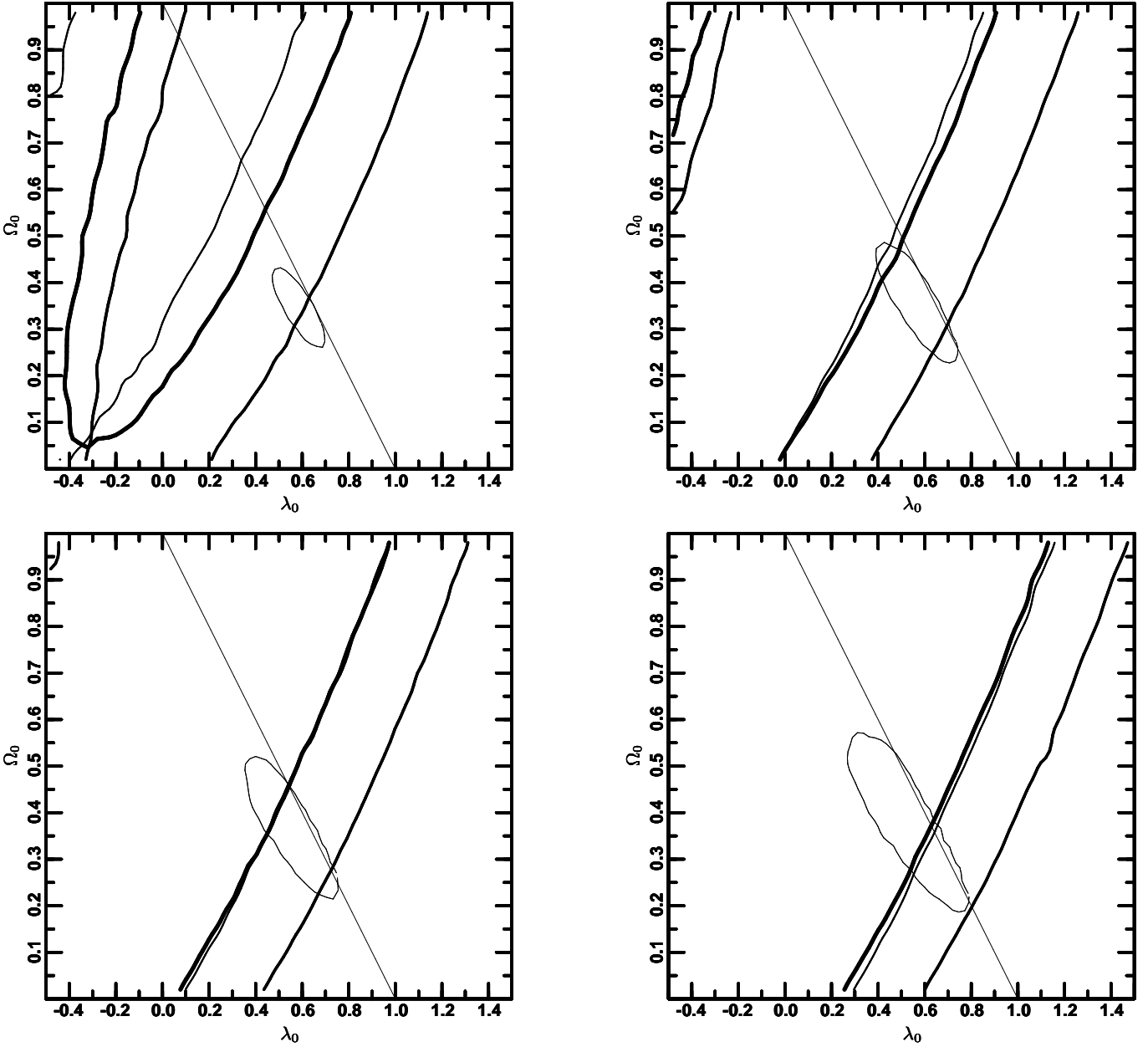

\noindent
\resizebox{0.45\textwidth}{!}{\includegraphics{8913.f05}}%
\makebox[0pt][r]{\resizebox{0.45\textwidth}{!}{\includegraphics{8913.f00}}}
\hfill
\resizebox{0.45\textwidth}{!}{\includegraphics{8913.f06}}%
\makebox[0pt][r]{\resizebox{0.45\textwidth}{!}{\includegraphics{8913.f00}}}

\vspace{1.5ex}

\noindent
\resizebox{0.45\textwidth}{!}{\includegraphics{8913.f07}}%
\makebox[0pt][r]{\resizebox{0.45\textwidth}{!}{\includegraphics{8913.f00}}}
\hfill
\resizebox{0.45\textwidth}{!}{\includegraphics{8913.f08}}%
\makebox[0pt][r]{\resizebox{0.45\textwidth}{!}{\includegraphics{8913.f00}}}
\caption[]{The 68\% (top left), 90\% (top right), 95\% (bottom left) and
99\% (bottom right) confidence contours for each of the data sets shown 
in Figs.~\ref{fi:cblik} and \ref{fi:lenses}.  In order of increasing 
thickness, the curves correspond to Fig.~\ref{fi:cblik} and, from left to 
right, the plots in Fig.~\ref{fi:lenses}}
\label{fi:overlap}
\end{figure*}

\begin{figure*}
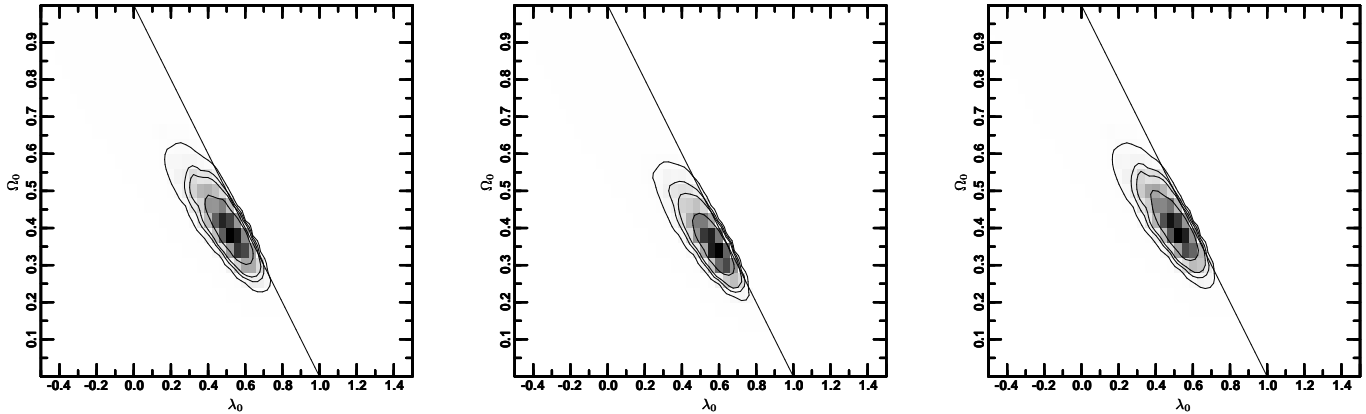

\noindent
\resizebox{0.3\textwidth}{!}{\includegraphics{8913.f09}}%
\makebox[0pt][r]{\resizebox{0.3\textwidth}{!}{\includegraphics{8913.f00}}}
\hfill
\resizebox{0.3\textwidth}{!}{\includegraphics{8913.f10}}%
\makebox[0pt][r]{\resizebox{0.3\textwidth}{!}{\includegraphics{8913.f00}}}
\hfill
\resizebox{0.3\textwidth}{!}{\includegraphics{8913.f11}}%
\makebox[0pt][r]{\resizebox{0.3\textwidth}{!}{\includegraphics{8913.f00}}}
\caption[]{
The same as Fig.~\ref{fi:lenses} but combined with the CMB constraints 
from Fig.~\ref{fi:cblik}
}
\label{fi:joint}
\end{figure*}

We follow the procedure outlined in Paper~III in computing joint 
constraints.  First, to make sure that the cosmological tests are not 
inconsistent with each other, we plot the overlap of various contours in 
Fig.~\ref{fi:overlap}.
%%%%%%%%%%%%%%%%%%%%%%%%%%%%%%%%%%%%%%%%%%%%%%%%%%%%%%%%%%%%%%%%%%%%%%%%
%%%%%%%%%%%%%%%%%%%%%%%%%%%%%%%%%%%%%%%%%%%%%%%%%%%%%%%%%%%%%%%%%%%%%%%%
%%%%%%%%%%%%%%%%%%%%%%%%%%%%%%%%%%%%%%%%%%%%%%%%%%%%%%%%%%%%%%%%%%%%%%%
In Fig.~\ref{fi:joint}, we present joint constraints formed by the 
multiplication of the corresponding probability density functions.
%%%%%%%%%%%%%%%%%%%%%%%%%%%%%%%%%%%%%%%%%%%%%%%%%%%%%%%%%%%%%%%%%%%%%%%
%%%%%%%%%%%%%%%%%%%%%%%%%%%%%%%%%%%%%%%%%%%%%%%%%%%%%%%%%%%%%%%%%%%%%%%
%%%%%%%%%%%%%%%%%%%%%%%%%%%%%%%%%%%%%%%%%%%%%%%%%%%%%%%%%%%%%%%%%%%%%%%%
There is (at least a small) region of overlap between the CMB contour 
and \emph{all} the lensing statistics contours at 90\% confidence (and of 
course at 95\% and 99\% as well), and with the JVAS contour (which we 
consider to be more reliable than the optical or joint contours) at 68\% 
confidence.  Thus, we consider the CMB and lensing constraints to be 
consistent or, at worse, only marginally inconsistent.

Note that the joint constraints using
lensing statistics and the CMB data differ only slightly from those
using the CMB data alone.  (It should be remembered that both the CMB and
the lensing constraints are probably too tight since all parameters
(except of course $\lambda_{0}$ and $\Omega_{0}$) have been fixed for
this analysis.  With more and better data, both can be expected to
improve in the future, while improvements in the theoretical models will
reduce systematic effects. However, since the input parameters to the
lensing statistics analysis are in many cases better understood than
those for the CMB analysis, the lensing statistics constraints are
probably more realistic than those from the CMB.  One should thus not
conclude that the CMB constraints make lensing statistics superfluous.)  
Nevertheless, the addition of even the current lens statistics data tightens the
upper limit on $\lambda_{0}$ and the lower limit on $\Omega_{0}$.  While
the CMB data alone provide perhaps the tightest constraints (with the
above-mentioned caveats) of any cosmological test, they still allow an
area of parameter space which is ruled out by other cosmological tests,
among which are lensing statistics.  Not only the upper (lower) limit on
$\lambda_{0}$ ($\Omega_{0}$) is tightened by adding lens statistics
constraints to those from the CMB, but also the best-fit cosmological
model shifts to a lower $\lambda_{0}$ and higher $\Omega_{0}$ value. 

The degeneracy in the $\lambda_{0}$-$\Omega_{0}$ plane is such that, in 
the region of non-negligible likelihood, the constraints from the CMB 
alone as well as the joint constraints with lensing statistics measure 
approximately $\lambda_{0} + \Omega_{0}$.  This region is described by 
the 99\% confidence contour, which covers the range in $\lambda_{0}$
of 0.3--0.8 for the CMB alone.  In the case of joint constraints, the 
region is shifted to lower $\lambda_{0}$ values as well as slightly 
increased in size, the exact values depending on the (combination of) 
lensing constraints used.
The corresponding range for $\Omega_{0}$ is 0.18--0.57, with a similar 
shifting to higher $\Omega_{0}$ values (and increase in range) for the 
joint constraints.
(Of course, the 99\% confidence contour is smaller than the 
rectangle defined by the ranges of
$\lambda_{0}$ and $\Omega_{0}$.)

Taken together, present measurements of cosmological parameters
\emph{definitely} rule out the Einstein-de~Sitter universe
($\lambda_{0}=0$, $\Omega_{0}=1$), \emph{very probably} rule out a
universe without a cosmological constant ($\lambda_{0}=0$) and
\emph{tentatively} rule out a flat ($\lambda_{0}$ + $\Omega_{0}$ = 1)
universe as well.\footnote{This \emph{tentative} conclusion should be 
considered with the necessary caution.  Apart from caveats arising from 
the limited parameter space explored (i.e.~all nuisance parameters were 
fixed), the confidence contours cannot be interpreted straightforwardly 
due to the fact that no closed models were computed.}
A universe with $\lambda_{0}\approx 0.4$ and
$\Omega_{0}\approx 0.3$ seems to be consistent with all observational
data, including measurements of the Hubble constant and age of the
universe.  It should be noted that it is really only the CMB data which
are indicating a possibly non-flat universe.  Other combinations of
cosmological tests \citep[e.g.][and references
therein]{MRoosSHoR99a,MTurner99a} tend to allow a flat universe within
the errors. 

\begin{figure}
\psfrag{l}{$l$}
\resizebox{\columnwidth}{!}{\includegraphics{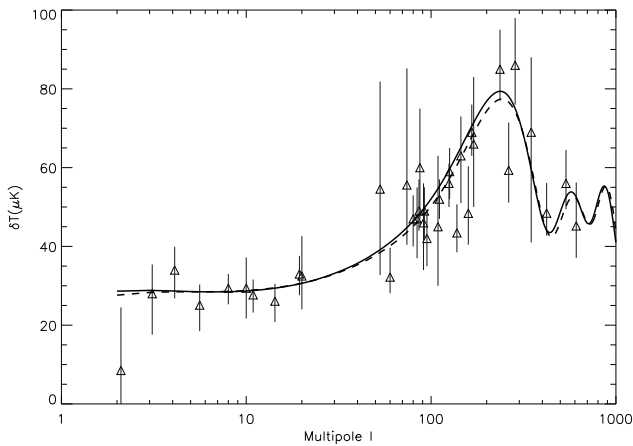}}
\caption[]{Data points with error bars and the power spectrum for our
best-fit model based on the CMB data alone or, as the values are the 
same, on the joint constraints from 
CMB and JVAS (solid) and for the combination of the CMB data with either 
the optical surveys discussed in Paper~I or with both the optical 
surveys and JVAS (dashed) (again, the values are the same)}
\label{fi:bestfit}
\end{figure}

The $\chi^{2}$ minimum for the CMB data is obtained for $\Omega_{0}=0.34$
and $\lambda_{0}= 0.60$.  The power spectrum (with the data points) for
this best-fit model is shown in the solid curve in Fig.~\ref{fi:bestfit}. 
This is also the best-fit model when the CMB constraints are combined 
with those from JVAS as in the centre plot of Fig.~\ref{fi:joint}.  The 
best-fit model for the combination of CMB and optical lensing 
constraints, either with or without the addition of JVAS constraints, 
has $\Omega_{0}=0.38$ and $\lambda_{0}= 0.52$; this is shown in the 
dashed curve of Fig.~\ref{fi:bestfit}.
%\suppressfloats
%%%%%%%%%%%%%%%%%%%%%%%%%%%%%%%%%%%%%%%%%%%%%%%%%%%%%%%%%%%%%%%%%%%%%%%%%%%
%%%%%%%%%%%%%%%%%%%%%%%%%%%%%%%%%%%%%%%%%%%%%%%%%%%%%%%%%%%%%%%%%%%%%%%%%%%
%%%%%%%%%%%%%%%%%%%%%%%%%%%%%%%%%%%%%%%%%%%%%%%%%%%%%%%%%%%%%%%%%%%%%%%%%%%

The comparison values from this work corresponding to those in
Table~3 of Paper~I are presented in Table~\ref{ta:special}.
%%%%%%%%%%%%%%%%%%%%%%%%%%%%%%%%%%%%%%%%%%%%%%%%%%%%%%%%%%%%%%%%%%%%%%%%%%%
\begin{table*}
\caption[]{Mean values and ranges for assorted confidence levels for the
parameter $\lambda_{0}$ for our likelihoods from this work, for the
special case $\Omega_{0}=0.3$.  This should be compared to Table~3 in
Paper~I.  Only lower limits are given for the case of the CMB alone, as
the upper limits all lie in the $k=+1$ area of parameter space, which
was not examined.  The contours near the $k=0$ line are thus not `real'
and should be ignored. Since $k=+1$ was not examined, no values for the
$k=0$ case can be extracted, as was the case in the corresponding tables
in Papers~I and II.  However, the $k=0$ special case has been examined
in detail in L97.  $X$ denotes the fact that there is no intersection 
of the confidence contour with the $\Omega_{0}=0.3$ line; equal upper and lower 
limits indicate a tangency} 
\label{ta:special}
\begin{tabular*}{\linewidth}{@{\extracolsep{\fill}}lrrrrrrrr}
\hline
Cosmological test&
\multicolumn{2}{c}{68\% c.l. range}  &
\multicolumn{2}{c}{90\% c.l. range}  &
\multicolumn{2}{c}{95\% c.l. range}  &
\multicolumn{2}{c}{99\% c.l. range}  \\
\hline
CMB, $p(D|\lambda_0)$ &
$0.58$ & $$ & $0.54$ & $$ & $0.52$ & $$ & $0.49$ & $$ \\
CMB \& optical, $p(D|\lambda_0)$ &
$0.62$ & $0.62$ & $0.54$ & $0.69$ & $0.52$ & $0.70$ & $0.49$ & $0.72$ \\
CMB \& JVAS, $p(D|\lambda_0)$ &
$0.58$ & $0.69$ & $0.54$ & $0.70$ & $0.52$ & $0.71$ & $0.49$ & $0.72$ \\
CMB \& optical \& JVAS, $p(D|\lambda_0)$ &
$X$ & $X$ & $0.55$ & $0.66$ & $0.52$ & $0.69$ & $0.50$ & $0.71$ \\
\hline
\end{tabular*}
\end{table*}
%%%%%%%%%%%%%%%%%%

Holding most of the parameters constant is of course a weak point of our
approach. Obviously, the goal is to explore the entire range of
parameter space, incorporating uncertainties in all parameters, prior
information etc. This is computationally very expensive.  Alternatively,
we could also think of a multiparameter maximisation method which would
provide a `best fit' value for all parameters, although assigning an
error would not be straightforward.  This might be risky because of
possible secondary maxima.  In fact, our calculations \emph{do} show a
secondary maximum, as can be seen by examining the data mentioned in the
URL below, although it is too weak to show up in the plots.  
The fact that the
secondary maximum occurs around $\lambda_{0}=-1$ and $\Omega_{0}=1$
looks suspicious, but there is nothing obviously wrong with the
behaviour of CMBFAST here (M.~Zaldarriaga, private
communication).\footnote{Initially, we \emph{did} find a bug in CMBFAST
for $\Omega_{0}=1$ and $\lambda_{0}<0$, but this has since been
corrected.}  This does appear to be `real' in the sense that it is what
the comparison of the data with the CMBFAST power spectra indicate.  Of
course, there might be unknown systematic effects in the former, but as
far as we can tell, there are no problems with CMBFAST which could
produce this.  On the other hand, it is probably not `real' in the sense
that it might disappear when more and/or better input data are used or
when a more exact code than the current version of CMBFAST---especially
for the relatively poorly explored area of parameter space where this
secondary maximum occurs---is used.  It should be kept in mind, however,
that there is no a priori reason to exclude a secondary maximum.  Also,
this secondary maximum appears in a part of parameter space which is
ruled out when a few cosmological tests are considered simultaneously
(see Paper~I), so in that sense it is also probably not `real'. 

The CMB data alone do not rule out closed ($k=+1$) 
models \citep[see also][]{MWhiteDScott96a}.  The
probability contours are thus compressed near the line that separates
the open models from the closed ones.  This is due to CMBFAST not (yet)
being able to make calculations for $k=+1$ models.  Even if these can be
ruled out by (some combination of) other cosmological tests, it would be
useful to extend the calculations formed here to include closed models,
which would allow for an easier interpretation of joint cosmological
constraints which include those from CMB data. 

As mentioned in Papers~I--III, to aid comparisons with other
cosmological tests, the data for the figures shown in this paper are
available at 
\begin{quote}
\texttt{http://multivac.jb.man.ac.uk:8000}\\
            \texttt{/ceres/data\symbol{95}from\symbol{95}papers/}
\end{quote}
or
\begin{quote}
\texttt{http://gladia.astro.rug.nl:8000}\\
            \texttt{/ceres/data\symbol{95}from\symbol{95}papers/}
\end{quote}
and we urge our colleagues to follow our example.

%%%%%%%%%%%%%%%%%%%%%%%%%%%%%%%%%%%%%%%%%%%%%%%%%%%%%%%%%%%%%%%%%%%%%%%%
\section{Conclusions}
\label{conclusions}

We have performed an analysis similar to that of \citet{CLineweaver98a},
but have used slightly different input data and a slightly different
statistical technique.  We have then combined the constraints in the
$\lambda_{0}$ $\Omega_{0}$ derived from the CMB with the results of the
constraints from gravitational lensing statistics presented in
\citet[Paper~I]{RQuastPHelbig99a} and \citet[Paper~II]{PHelbigMQWBK99a}.

Using the CMB data alone, the best-fit model has $\lambda_{0}=0.6$ and
$\Omega_{0}=0.34$ and at 95\% confidence the lower limit on
$\lambda_{0}+\Omega_{0}$ is $0.8$.  Including constraints from
gravitational lensing statistics doesn't change this significantly, 
although it does change the allowed region of parameter space. 
A universe with
$\lambda_{0}=0$ is ruled out for any value of $\Omega_{0}$ at better
than 99\% confidence using the CMB alone. Combined with constraints from
lensing statistics, $\lambda_{0}=0$ is also ruled out 
at better than 99\% confidence. 

As the region of parameter space allowed by the CMB is, within our 
assumptions, much smaller than that allowed by lensing statistics, the 
main result of combining the two is to change the range of parameter 
space allowed by the CMB along its axis of degeneracy.  This axis of 
degeneracy is along a line of constant $\Omega_{0}+\lambda_{0}$, 
i.e.~along a line of constant radius of curvature.  Indeed, it is close 
to the $\Omega_{0}+\lambda_{0}=1$ line, which corresponds to a flat 
universe. 

The CMB analysis favours $\lambda_{0}\approx0.60$ and
$\Omega_{0}\approx0.34$.  This confirms the long-established result that
the $\lambda_{0}=0$, $\Omega_{0}=1$ (Einstein-de~Sitter) model is ruled
out by the data and supports the recent evidence for a positive
cosmological constant (see Papers~II and III for a discussion).  The 
combination
of results discussed in Paper~II and \citet{CLineweaver98a} hinted
tentatively that flat ($k=0$) cosmological models are beginning to be
ruled out by the data.  However, to quantify this, it would be
interesting to compute power spectra for spatially closed ($k=+1$)
cosmological models, since these are not ruled out by current CMB data. 

One should keep in mind that these conclusions assume fixed values for
the nuisance parameters.  On the other hand, these fixed values
correspond to values which are (currently) generally accepted and/or
values at the global $\chi^{2}$ minimum in a larger parameter space
examined in L98. 

On the one hand, the CMB data provide good evidence for a flat universe,
since the allowed region of parameter space is very small and very near
the $k=0$ line.  On the other hand, as noted in L98, the allowed region 
is so small that a significant departure from $k=0$ is hinted at.  
(However, one should keep in mind that either of these might be a 
consequence of not taking the uncertainties in the other parameters 
fully into account.)  Nevertheless, compared to other cosmological tests 
which allow a larger portion of the $\lambda_{0}$-$\Omega_{0}$ plane, 
the CMB provides a strong hint that the universe is close to being flat.
On the other hand, the \emph{combination} of the CMB data and the data 
from other cosmological tests tend to indicate that the best fit might 
actually be achieved for $k<0$, as discussed in Paper~I.  While various 
tests might individually allow $k=0$, they do so for different values of 
$\lambda_{0}$ (or, equivalently, $\Omega_{0}$ which of course is
$1-\lambda_{0}$ in a flat universe) so that, in combination, they 
provide evidence against a flat universe.

If the universe does have $k=0$ or is arbitrarily close to it, this can
never be proven in practice, though our confidence in a measurement
indicating $k=0$ would be inversely proportional to the size of the
error bars.  On the other hand, if the universe is in fact not flat,
then this \emph{can} be proven, by reducing the error bars so that the
$k=0$ case is ruled out.  At present, the question of the sign of $k$ or 
equivalently (assuming a simple topology) the question whether the universe 
is finite or infinite is still an open question.  On the other hand, 
there is strong evidence for the fact that it will expand forever.

Assuming the universe is not flat then, since it is relatively close to
being so, to demonstrate that it is not flat it will be necessary to
reduce the systematic errors in the comparison of observations with
theory. This can be done by incorporating the uncertainties in all input
parameters into the calculations (for CMB constraints and other
cosmological tests).  Of course, it is also necessary to explore a large
enough region of parameter space, in all dimensions, in sufficient
resolution. 

While the joint constraints leave only a small area of the
$\lambda_{0}$-$\Omega_{0}$ parameter space which fits the observations,
it should be remembered that neither the CMB nor the lens statitistics
analyses we have performed incorporates uncertainties in the input
parameters: all parameters except $\lambda_{0}$ and $\Omega_{0}$ have
been held constant.  This is sensible in a first-step approach, but of
course an analysis of the full parameter space should be performed in
order to get robust constraints on $\lambda_{0}$, $\Omega_{0}$ and the
other parameters these analyses are sensitive to.  Our suspicion is that
as a result of this the CMB constraints will relax more than the lensing
statistics constraints, so, despite the impression that the figures here
give that the CMB on its own is powerful enough to constrain the
cosmological parameters, one should also include gravitational lensing
statistics in `joint constraints' analyses
\citep[e.g.][]{MWhite98a,DEisensteinHT98b,MTegmarkEH98a,MTegmarkEHK98a,
DEisensteinHT98a,MWebsterBHLLR98a}.  This will become more important in 
the future with the completion of large, well-defined gravitational lens 
surveys such as CLASS \citep[e.g.][]{SMyersetal99a}.

\begin{acknowledgements}
It is a pleasure to thank D.\ Barbosa, E.\ Bunn, G.\ Hinshaw and
C.\ Lineweaver for helpful discussions and M.\ Zaldarriaga and U.\ Seljak
for making their CMBFAST code publicly available.  This research was
supported in part by the European Commission, TMR Programme, Research
Network Contract ERBFMRXCT96-0034 `CERES'.  JFMP acknowledges the
support of a PPARC studentship. 
\end{acknowledgements}

\end{document}